\documentclass[11pt,a4paper]{article}
\pdfoutput=1
\usepackage{jheppub}
\usepackage{tikz}
\usetikzlibrary{intersections, calc, arrows.meta}
\usepackage{subcaption}
\usepackage{simpler-wick}
\usepackage{bm}

\DeclareMathOperator{\tr}{tr}

\newcommand{\ri}{\mathrm{i}}
\renewcommand{\th}{\theta}

\newcommand{\cob}{\delta}

\newcommand{\hf}{\frac{1}{2}}

\renewcommand{\b}[1]{\overline{#1}}
\newcommand{\del}{\partial}

\newcommand{\lap}{\Delta}
\newcommand{\bra}{\langle}
\newcommand{\ket}{\rangle}
\newcommand{\la}{\lambda}

\newcommand{\h}[1]{\widehat{#1}}
\newcommand{\bt}{\beta}

\newcommand{\rt}[1]{\sqrt{#1}}

\newcommand{\cC}{\mathcal{C}}

\newcommand{\mZ}{\mathbb{Z}}

\newcommand{\qb}[1]{\Bigl[~\begin{matrix}#1\end{matrix}~\Bigr]_q}

\begin{document}

\title{Hartle-Hawking wavefunction in double scaled SYK}

\author{Kazumi Okuyama}

\affiliation{Department of Physics, 
Shinshu University, 3-1-1 Asahi, Matsumoto 390-8621, Japan}

\emailAdd{kazumi@azusa.shinshu-u.ac.jp}

\abstract{
We compute the transition amplitude between the chord number $0$ and $\ell$ states in the double scaled SYK model and interpret it as a Hartle-Hawking wavefunction of the bulk gravitational theory. We observe that the so-called un-crossed matter correlators of double scaled SYK model are obtained by gluing the Hartle-Hawking wavefunctions with an appropriate weight.
}

\maketitle

\section{Introduction}
Sachdev-Ye-Kitaev (SYK) model \cite{kitaev1,kitaev2,Sachdev1993,Maldacena:2016hyu,Polchinski:2016xgd}
is a useful toy model for the study of quantum
gravity.
The low energy sector of SYK model
is described by the Schwarzian mode and this part of the
dynamics is holographically dual to Jackiw-Teitelboim (JT)
gravity \cite{Jackiw:1984je,Teitelboim:1983ux}.
As discussed in \cite{Cotler:2016fpe,Berkooz:2018qkz,Berkooz:2018jqr},
one can go beyond this low energy approximation by taking
a certain double scaling limit of the SYK model.
It turns out that this double scaled SYK (DSSYK) model
is exactly solvable using the 
technique of the chord diagram 
\cite{Berkooz:2018qkz,Berkooz:2018jqr}.
However, as emphasized in \cite{Lin:2022rbf}, 
the bulk gravitational picture of DSSYK
is not well understood.
One mysterious feature of DSSYK is that
the geodesic length of bulk spacetime becomes a discrete chord number
$\ell$.
It is proposed in \cite{Lin:2022rbf} that
the chord number state $|\ell\ket$ originally introduced
in \cite{Berkooz:2018jqr} should be identified as the 
state of bulk Hilbert space
and the expression of the partition function
$Z(\bt)=\bra0|e^{-\bt T}|0\ket$ in terms of the
transfer matrix $T$ acting on the chord number states has a 
natural bulk gravitational interpretation. 

Generalizing this interpretation in \cite{Lin:2022rbf},
we propose that the amplitude
$\bra\ell|e^{-\bt T}|0\ket$ is interpreted as the Hartle-Hawking 
(HH) wavefunction
of the bulk gravitational theory \cite{Hartle:1983ai}
and we explicitly compute $\bra\ell|e^{-\bt T}|0\ket$.
We realize that
the HH wavefunction has been secretly appeared in 
appendix C of \cite{Berkooz:2018jqr}
in the computation of the so-called un-crossed
matter correlators in DSSYK.
Thus we find a bulk gravitational interpretation
of the un-crossed correlators: these correlators are obtained by
gluing the HH wavefunctions with an appropriate weight.

This paper is organized as follows. In section \ref{sec:review}, 
we review the solution of DSSYK in terms of the chord diagrams and
the transfer matrix \cite{Berkooz:2018jqr}.
In section
\ref{sec:HH}, we compute the HH wavefunction 
$\bra\ell|e^{-\bt T}|0\ket$.
In section \ref{sec:matter}, we present our observation 
that the un-crossed correlators in appendix C of 
\cite{Berkooz:2018jqr} are written in terms of the HH wavefunctions.
In section \ref{sec:prob}, we consider the
probability distribution 
of chord number
$\ell$ given by the absolute value squared 
$|\bra\ell|e^{-\bt T}|0\ket|^2$
of the HH wavefunction. 
Finally we conclude in section 
\ref{sec:conclusion} with some discussion on the future problems.

\section{Review of doubled scaled SYK}\label{sec:review}
We first review the result of DSSYK in \cite{Berkooz:2018jqr}.
SYK model is defined by the Hamiltonian for 
$N$ Majorana fermions $\psi_i~(i=1,\cdots,N)$
with all-to-all $p$-body interaction
\begin{equation}
\begin{aligned}
 H=\ri^{p/2}\sum_{i_1,\cdots,i_p=1}^NJ_{i_1\cdots i_p}\psi_{i_1}\cdots\psi_{i_p},
\end{aligned} 
\end{equation}
where $\psi_i$'s obey the anti-commutation relation $\{\psi_i,\psi_j\}=2\cob_{i,j}$
and the random coupling $J_{i_1\cdots i_p}$ is drawn from
the Gaussian distribution 
\begin{equation}
\begin{aligned}
 \bra J_{i_1\cdots i_p}\ket_J=0,\quad
\bra J_{i_1\cdots i_p}^2\ket_J=\binom{N}{p}^{-1}.
\end{aligned} 
\end{equation}
DSSYK is defined by the limit
\begin{equation}
\begin{aligned}
 N,p\to\infty\quad\text{with}\quad\la=\frac{2p^2}{N}:\text{fixed}.
\end{aligned} 
\label{eq:scaling}
\end{equation}
As shown in \cite{Berkooz:2018jqr}, SYK model is exactly solvable in this double scaling limit.
In the computation of the trace $\bra \tr H^k\ket_J$,
the average over the random coupling
$J$ can be organized
as the so-called chord diagram arising from the Wick contraction of
$J$'s and the remaining trace of Majorana fermions
reduces to a counting problem
of the intersection number of chords
\begin{equation}
\begin{aligned}
 \bra \tr H^k\ket_J=\sum_{\text{chord diagrams}}q^{\#(\text{intersections})},
\end{aligned} 
\end{equation}
where $q$ is given by
\begin{equation}
\begin{aligned}
 q=e^{-\la}.
\end{aligned} 
\end{equation}
This counting problem is solved by introducing the transfer matrix acting on the
states $\{|n\ket\}_{n=0,1,\cdots}$. Here $|n\ket$ denotes
the state with $n$ chords 
\begin{equation}
\begin{aligned}
 \begin{tikzpicture}[scale=0.75]
\draw (0,-1)--(0,1);
\draw (0.5,-1)--(0.5,1);
\draw (1,-1)--(1,1);
\draw (1.5,-1)--(1.5,1);
\draw (2,-1)--(2,1);
\draw (2.5,-1)--(2.5,1);
\draw (3,-1)--(3,1);
\draw[dashed] (-0.5,0)--(3.5,0);
\draw (-1,0) node [left]{$|n\ket=$};
\draw (1.5,1) node [above] {$\overbrace{~\hskip23mm~}^{}$};
\draw (1.5,1.5) node [above] {$n~\text{chords}$};
\end{tikzpicture}
\end{aligned} 
\label{eq:def-n}
\end{equation}
and the dashed line in \eqref{eq:def-n} represents
a constant (Euclidean) time slice. We normalize $|n\ket$ as
\begin{equation}
\begin{aligned}
 \bra n|m\ket=\cob_{n,m},\quad
1=\sum_{n=0}^\infty |n\ket\bra n|.
\end{aligned} 
\end{equation}
Then the partition function $Z(\bt)$ of DSSYK is written as
\begin{equation}
\begin{aligned}
 Z(\bt)=\bra \tr e^{-\bt H}\ket_J=\bra 0|e^{-\bt T}|0\ket,
\end{aligned} 
\label{eq:Z}
\end{equation}
where the transfer matrix $T$ is written in terms
of the $q$-deformed oscillator $A_{\pm}$ as
\begin{equation}
\begin{aligned}
 -T=A_-+A_+.
\end{aligned} 
\label{eq:T-def}
\end{equation}
$A_{\pm}$ acts on the state $|n\ket$ as
\begin{equation}
\begin{aligned}
 A_-|n\ket=\rt{[n]}|n-1\ket,\quad A_+|n\ket=\rt{[n+1]}|n+1\ket,
\end{aligned} 
\label{eq:Apm-act}
\end{equation} 
where $[n]$ is the $q$-integer
\begin{equation}
\begin{aligned}
 {[}n{]}=\frac{1-q^n}{1-q}.
\end{aligned} 
\end{equation}
From \eqref{eq:Apm-act}, one can easily show that
$A_\pm$ obey the relations
\begin{equation}
\begin{aligned}
 A_-A_+-qA_+A_-=1,\quad [A_-,A_+]=q^{\h{N}},
\end{aligned} 
\end{equation}
where $\h{N}$ is the number operator
\begin{equation}
\begin{aligned}
 \h{N}|n\ket=n|n\ket.
\end{aligned} 
\label{eq:hat-N}
\end{equation}
Some comments are in order here:
\begin{enumerate}
 \item Our $-T$ in \eqref{eq:T-def} is equal to $\hat{T}$ in 
\cite{Berkooz:2018jqr}, which is related to the original $T$ in 
\cite{Berkooz:2018jqr} by a conjugation.
Put differently, the original $T$ in 
\cite{Berkooz:2018jqr} is obtained from our $-T$ in \eqref{eq:T-def}
by acting it on a different basis $\{A_+^n|0\ket\}_{n=0,1,\cdots}$
\begin{equation}
\begin{aligned}
 -TA_+^n|0\ket=A_+^{n+1}|0\ket+\frac{1-q^n}{1-q}A_+^{n-1}|0\ket,
\end{aligned} 
\end{equation}
where $-T=A_-+A_+$ in \eqref{eq:T-def}. Note that the state $A_+^n|0\ket$
is related to $|n\ket$ as
\begin{equation}
\begin{aligned}
 A_+^n|0\ket=\rt{[n]!}|n\ket,\qquad [n]!=\prod_{k=1}^n [k],
\end{aligned} 
\end{equation}
and the norm of $A_+^n|0\ket$ is given by the $q$-factorial
of $n$
\begin{equation}
\begin{aligned}
 \big|A_+^n|0\ket\big|^2=[n]!.
\end{aligned} 
\end{equation} 
This basis and normalization were adopted in \cite{Lin:2022rbf}.
\item We have put the minus sign in the definition of $T$ in \eqref{eq:T-def}.
This minus sign does not change the final result since the spectrum of the Hamiltonian
is symmetric under the sign flip $H\to -H$.
This symmetry can be traced back to the Gaussian distribution
of the random coupling $J$, where $+J$ and $-J$ appear with an equal probability. 
\end{enumerate}

As shown in \cite{Berkooz:2018jqr}, one can compute the partition function
in \eqref{eq:Z} by going to the eigenbasis of the transfer matrix $T$, 
which is given by the $q$-Hermite polynomial
$H_n(\cos\th|q)$
\begin{equation}
\begin{aligned}
 \bra \th|n\ket=\frac{H_n(\cos\th|q)}{\rt{(q;q)_n}},
\end{aligned} 
\end{equation} 
where $(a;q)_n$ denotes the $q$-Pochhammer symbol
\begin{equation}
\begin{aligned}
 (a;q)_n=\prod_{k=0}^{n-1}(1-aq^k),
\end{aligned} 
\label{eq:q-poch}
\end{equation}
and the $q$-Hermite polynomial is defined by
\begin{equation}
\begin{aligned}
 H_n(\cos\th|q)=\sum_{k=0}^n\qb{n\\k}e^{\ri(n-2k)\th},
\end{aligned} 
\label{eq:q-hermite}
\end{equation}
with $\qb{n\\k}$ being the $q$-binomial coefficient 
\begin{equation}
\begin{aligned}
 \qb{n\\k}=\frac{(q;q)_n}{(q;q)_k(q;q)_{n-k}}.
\end{aligned} 
\label{eq:q-binom}
\end{equation}

Using the recurrence relation for the
$q$-Hermite polynomial, one can show that $T$ is diagonal in this basis
\begin{equation}
\begin{aligned}
 \bra \th|T|n\ket=E(\th)\bra \th|n\ket,
\end{aligned} 
\end{equation}
and the eigenvalue $E(\th)$ is given by
\begin{equation}
\begin{aligned}
 E(\th)=-\frac{2\cos\th}{\rt{1-q}}.
\end{aligned} 
\end{equation}
$\bra\th|n\ket$ obeys an orthogonality relation with respect to a
certain
integration measure of $\th$
\begin{equation}
\begin{aligned}
 \int_0^\pi\frac{d\th}{2\pi}\mu(\th)\bra n|\th\ket\bra \th|m\ket=\cob_{n,m},
\end{aligned} 
\end{equation}
where the measure factor $\mu(\th)$ is given by
\begin{equation}
\begin{aligned}
 \mu(\th)=(q;q)_\infty (e^{2\ri\th};q)_\infty(e^{-2\ri\th};q)_\infty.
\end{aligned} 
\end{equation}
Then the partition function \eqref{eq:Z} is written as
\begin{equation}
\begin{aligned}
 \bra 0|e^{-\bt T}|0\ket&=\int_0^\pi\frac{d\th}{2\pi}\mu(\th) \bra 0|\th\ket e^{-\bt E(\th)}
\bra \th|0\ket\\
&=\int_0^\pi\frac{d\th}{2\pi}\mu(\th)e^{-\bt E(\th)},
\end{aligned} 
\end{equation}
where we used $\bra\th|0\ket=\frac{H_0(\cos\th|q)}{\rt{(q;q)_0}}=1$.
This integral was evaluated in \cite{Berkooz:2018jqr} as
\begin{equation}
\begin{aligned}
\bra 0|e^{-\bt T}|0\ket
=\sum_{r=0}^\infty (-1)^r q^{\hf r(r+1)}(2r+1)\frac{\rt{1-q}}{\bt}I_{2r+1}
\left(\frac{2\bt}{\rt{1-q}}\right),
\end{aligned} 
\end{equation}
where $I_n(z)$ denotes the modified Bessel function
of the first kind. In the next section, we will compute the more general 
amplitude of $e^{-\bt T}$ sandwiched between $n=0$ and $n=\ell$
\begin{equation}
\begin{aligned}
 \bra \ell|e^{-\bt T}|0\ket,
\end{aligned} 
\label{eq:l-0}
\end{equation}
which we interpret as the Hartle-Hawking (HH) wavefunction of the bulk 
gravitational theory.

\section{Hartle-Hawking wavefunction in doubled scaled SYK}\label{sec:HH}
Let us consider the amplitude $\bra \ell|e^{-\bt T}|0\ket$
in \eqref{eq:l-0}.
In the bulk gravitational picture, it is suggested in \cite{Berkooz:2018qkz,Berkooz:2018jqr,Lin:2022rbf} that
the chord number $\ell$
in $|\ell\ket$ can be interpreted as the 
discretized version of the geodesic length
of the bulk spacetime.
The bulk picture
of the amplitude $\bra \ell|e^{-\bt T}|0\ket$
is depicted as
\begin{equation}
\begin{aligned}
\begin{tikzpicture}[scale=0.75]
\draw (-2,0) arc [start angle=-180,end angle=0,radius=2];
\draw(-2,0)--(2,0); 
\draw (0,0) node [above]{$\ell$};
\draw (0,-2) node [below]{$\bt$};
\draw (-2.5,-0.5) node [left]{$\bra \ell|e^{-\bt T}|0\ket=$};
\end{tikzpicture}
\end{aligned}
\label{eq:HH-pic} 
\end{equation}
which is naturally interpreted as the HH wavefunction of
the bulk gravitational theory. According to the rule in \eqref{eq:def-n},
$\ell$ chords are threading the top horizontal line of \eqref{eq:HH-pic}. 
As we will see in the next section,
this HH wavefunction $\bra \ell|e^{-\bt T}|0\ket$ serves as a basic building block
for the correlation functions of matter fields. 

The HH wavefunction $\bra \ell|e^{-\bt T}|0\ket$ can be computed by
going to the eigenbasis $|\th\ket$ of the transfer matrix
\begin{equation}
\begin{aligned}
 \bra \ell|e^{-\bt T}|0\ket&=\int_0^\pi\frac{d\th}{2\pi}\mu(\th)\bra \ell|\th\ket e^{-\bt E(\th)}\bra \th|0\ket\\
&=\int_0^\pi\frac{d\th}{2\pi}\mu(\th)\bra \ell|\th\ket e^{-\bt E(\th)},
\end{aligned} 
\label{eq:HH-int}
\end{equation}
where we used $\bra\th|0\ket=1$. In order to evaluate this integral,
we first note that the measure factor
$\mu(\th)$ is expanded as
\begin{equation}
\begin{aligned}
 \mu(\th)&=-(e^{\ri\th}-e^{-\ri\th})^2\prod_{n=1}^\infty (1-q^n)
(1-e^{2\ri \th}q^n)(1-e^{-2\ri \th}q^n)\\
&=-2\ri\sin\th\sum_{r\in\mZ}(-1)^r q^{\hf r(r+1)}e^{\ri(2r+1)\th}.
\end{aligned} 
\label{eq:mu-exp}
\end{equation}
Combining the first factor $-2\ri\sin\th$ of \eqref{eq:mu-exp}
and the Boltzmann weight $e^{-\bt E(\th)}$, we find
\begin{equation}
\begin{aligned}
 -2\ri\sin\th e^{-\bt E(\th)}&=\frac{\rt{1-q}}{\bt}\ri\del_\th e^{-\bt E(\th)}\\
&=\frac{\rt{1-q}}{\bt}\ri\del_\th\sum_{n\in\mZ}
I_n\left(\frac{2\bt}{\rt{1-q}}\right)e^{\ri n\th}\\
&=\frac{\rt{1-q}}{\bt}\sum_{n\in\mZ} (-n)I_n\left(\frac{2\bt}{\rt{1-q}}\right)
e^{\ri n\th}.
\end{aligned} 
\label{eq:bessel-exp}
\end{equation}
Now, using \eqref{eq:q-hermite}, \eqref{eq:mu-exp} and
\eqref{eq:bessel-exp}, the $\th$-integral
in \eqref{eq:HH-int} is evaluated as
\begin{equation}
\begin{aligned}
 \bra\ell|e^{-\bt T}|0\ket&=\frac{1}{\rt{(q;q)_\ell}}
\sum_{r\in\mZ}(-1)^r q^{\hf r(r+1)}\sum_{k=0}^\ell
\qb{\ell\\k}(2r+1+\ell-2k)
\frac{\rt{1-q}}{2\bt}I_{2r+1+\ell-2k}\left(\frac{2\bt}{\rt{1-q}}\right).
\end{aligned} 
\label{eq:Z-wa}
\end{equation}
After shifting $r\to r+k$, the sum over $k$ can be performed 
with the help of the $q$-binomial formula
\begin{equation}
\begin{aligned}
 (a;q)_n=\sum_{k=0}^n (-a)^k q^{\hf k(k-1)}\qb{n\\k}.
\end{aligned} 
\end{equation}
After some algebra we find
\begin{equation}
\begin{aligned}
 \bra\ell|e^{-\bt T}|0\ket&=\sum_{r\in\mZ}(-1)^r q^{\hf r(r+1)}
\frac{(q;q)_{r+\ell}}{\rt{(q;q)_\ell}(q;q)_r}(2r+1+\ell)
\frac{\rt{1-q}}{2\bt}I_{2r+1+\ell}\left(\frac{2\bt}{\rt{1-q}}\right).
\end{aligned} 
\label{eq:HH-sum0}
\end{equation}
One can further simplify this expression 
as follows.
Using the relation
\begin{equation}
\begin{aligned}
 \frac{(q;q)_{r+\ell}}{(q;q)_r}=0,\qquad(0>r\geq-\ell),
\end{aligned} 
\end{equation}
the summation in \eqref{eq:HH-sum0} can be divided into two parts
\begin{equation}
\begin{aligned}
 \bra\ell|e^{-\bt T}|0\ket&=\left(\sum_{r\geq0}+\sum_{r\leq-\ell-1}\right)(-1)^r q^{\hf r(r+1)}
\frac{(q;q)_{r+\ell}}{\rt{(q;q)_\ell}(q;q)_r}(2r+1+\ell)
\frac{\rt{1-q}}{2\bt}I_{2r+1+\ell}\left(\frac{2\bt}{\rt{1-q}}\right).
\end{aligned} 
\end{equation} 
One can show that the two contributions $\sum_{r\geq0}$
and $\sum_{r\leq-\ell-1}$
are related by the transformation $r\to-r-\ell-1$ and 
these two
contributions are actually equal.
Thus we can restrict the summation to $r\geq0$ and 
multiply by the factor of $2$.
Finally we find
\begin{equation}
\begin{aligned}
 \bra\ell|e^{-\bt T}|0\ket&=\sum_{r=0}^\infty(-1)^r q^{\hf r(r+1)}
\frac{(q;q)_{r+\ell}}{\rt{(q;q)_\ell}(q;q)_r}(2r+1+\ell)
\frac{\rt{1-q}}{\bt}I_{2r+1+\ell}\left(\frac{2\bt}{\rt{1-q}}\right).
\end{aligned} 
\label{eq:HH-sum}
\end{equation}
This is the main result of this section.
As a consistency check, one can see that
when $\ell=0$ this reduces to the known result
\eqref{eq:Z-wa} of partition function, as expected.

One can generalize this computation to the more general overlap
$\bra\ell_1|e^{-\bt T}|\ell_2\ket$, known as the propagator
(see \cite{Saad:2019pqd} for the propagator in JT gravity).
In a similar manner as above, we find
\begin{equation}
\begin{aligned}
 \bra\ell_1|e^{-\bt T}|\ell_2\ket
=\sum_{r\in\mZ}(-1)^r q^{\hf r(r+1)}
\frac{\cC_{\ell_1,\ell_2,r}}{\rt{(q;q)_{\ell_1}(q;q)_{\ell_2}}}(2r+1+\ell_1+\ell_2)
\frac{\rt{1-q}}{2\bt}I_{2r+1+\ell_1+\ell_2}\left(\frac{2\bt}{\rt{1-q}}\right),
\end{aligned} 
\end{equation}
where $\cC_{\ell_1,\ell_2,r}$ is given by
\begin{equation}
\begin{aligned}
 \cC_{\ell_1,\ell_2,r}=\sum_{k_1=0}^{\ell_1}
\sum_{k_2=0}^{\ell_2}(-1)^{k_1+k_2}\qb{\ell_1\\ k_1}
\qb{\ell_2\\ k_2}q^{\hf(k_1+k_2)(k_1+k_2+2r+1)}.
\end{aligned} 
\end{equation}
When $\ell_2=0$, $\cC_{\ell_1,\ell_2,r}$ becomes
\begin{equation}
\begin{aligned}
 \cC_{\ell,0,r}=\frac{(q;q)_{r+\ell}}{(q;q)_r},
\end{aligned} 
\end{equation}
which reproduces our result of the HH wavefunction in \eqref{eq:HH-sum0}.
We note in passing that the propagator
$\bra\ell_1|e^{-\bt T}|\ell_2\ket$ is symmetric
under the exchange of $\ell_1$ and
$\ell_2$
\begin{equation}
\begin{aligned}
 \bra\ell_1|e^{-\bt T}|\ell_2\ket=\bra\ell_2|e^{-\bt T}|\ell_1\ket.
\end{aligned} 
\end{equation}

\section{Un-crossed matter correlators}\label{sec:matter}
As discussed in \cite{Berkooz:2018jqr}, one can introduce
a matter field in the bulk which is dual to an
operator in the DSSYK model. One simple example 
is the length $s$ strings of Majorana fermions
\begin{equation}
\begin{aligned}
 M=\ri^{s/2}\sum_{i_1\cdots i_s}K_{i_1\cdots i_s}\psi_{i_1}\cdots\psi_{i_s}
\end{aligned} 
\end{equation}
with Gaussian random coefficients $K_{i_1\cdots i_s}$
which is drawn independently from the random coupling
$J$ in the SYK Hamiltonian.
The effect of this operator can be made finite by taking the scaling limit
\begin{equation}
\begin{aligned}
 N,p,s\to\infty\quad\text{with}\quad\lap=\frac{2ps}{N}:\text{fixed}.
\end{aligned} 
\end{equation}
In this limit, the random average of the correlator
such as $\tr(e^{-\bt_2 H}Me^{-\bt_1 H}M)$ can be computed 
using the technique of the chord diagram. Compared to the computation of the
partition function, one novel feature is that we have to introduce
a new type of chord coming from the Wick contraction of random variables 
$K_{i_1\cdots i_s}$, and the combinatorics is schematically written as
\begin{equation}
\begin{aligned}
 \bra\tr(e^{-\bt_2 H}\wick{\c Me^{-\bt_1 H}\c M})\ket_J
=\sum_{\text{chord diagrams}}q^{\# (H\text{-}H~\text{intersections})}e^{-\lap
\#(H\text{-}M~\text{intersections})}.
\end{aligned} 
\end{equation}
As shown in \cite{Berkooz:2018jqr}, if the 
chords coming from the matter operators do not intersect with each other, 
the correlator can be evaluated in a rather explicit form.
This type of correlator is called ``un-crossed correlator'' in
\cite{Berkooz:2018jqr}. In this section we consider the
un-crossed $2n$-point function
\begin{equation}
\begin{aligned}
 \bra\tr e^{-\bt_{2n}H}\wick{\c M_n e^{-\bt_{2n-1}H}\c M_n}\cdots
\wick {\c M_1e^{-\bt_1H}\c M_1}\ket_J.
\end{aligned} 
\end{equation}

\subsection{2-point function}
As a warm up, let us first consider the
2-point function of matter operators
\begin{equation}
\begin{aligned}
 \bra \tr(e^{-\bt_2 H}\wick{\c Me^{-\bt_1 H} \c M})\ket_J.
\end{aligned} 
\end{equation} 
In the transfer matrix formalism, the effect of 
the Wick contraction of matter is summarized as
\footnote{As shown in \cite{Berkooz:2018jqr}, 
the bi-local operator $\wick{\c M e^{-\bt H} \c M}$
is written as
\begin{equation}
\begin{aligned}
 \wick{\c Me^{-\bt H} \c M}=
\sum_{\ell,\ell'=0}^\infty \sum_{i=0}^{\min(\ell,\ell')}
|\ell')(\ell'-i|
e^{-\lap\h{N}} e^{-\bt T} e^{-\lap\h{N}} |\ell-i)(\ell|P^{(\ell)}_i,
\end{aligned}
\label{eq:biO} 
\end{equation}
where 
\begin{equation}
\begin{aligned}
 P^{(\ell)}_i=(e^{-2\lap};q)_i\qb{\ell\\ i},\quad
|\ell)=\rt{[\ell]!}|\ell\ket,\quad
(\ell|=\frac{1}{\rt{[\ell]!}}\bra \ell|.
\end{aligned} 
\end{equation}
One can check that \eqref{eq:biO} 
reduces to \eqref{eq:matter-wick} after shifting $\ell\to\ell+i,\ell'\to\ell'+i$.
Note that $P^{(\ell)}_i$ satisfies
\begin{equation}
\begin{aligned}
 \sum_{i=0}^\ell e^{-2\lap(\ell-i)}P^{(\ell)}_i=1,
\end{aligned} 
\end{equation}
which guarantees $\wick{\c M \c M}=1$, i.e., $\wick{\c M e^{-\bt H} \c M}$
becomes the identity operator when $\bt=0$.
}
\begin{equation}
\begin{aligned}
 \wick{\c Me^{-\bt H} \c M}=\sum_{\ell,\ell',i=0}^\infty 
(e^{-2\lap};q)_i\rt{\qb{\ell+i\\ i}\qb{\ell'+i\\ i}}~
|\ell'+i\ket\bra \ell'|e^{-\lap\h{N}} e^{-\bt T} e^{-\lap\h{N}} 
|\ell\ket\bra \ell+i|,
\end{aligned} 
\label{eq:matter-wick}
\end{equation}
where $\h{N}$ is the number operator defined in \eqref{eq:hat-N}. 
Then the 2-point function is written as
\begin{equation}
\begin{aligned}
& \bra \tr(e^{-\bt_2 H}\wick{\c Me^{-\bt_1 H} \c M})\ket_J\\
=&\bra 0|
e^{-\bt_2 T}\wick{\c Me^{-\bt_1 H} \c M}|0\ket\\
=&\bra 0|e^{-\bt_2 T}e^{-\lap \h{N}}e^{-\bt_1 T}|0\ket\\
=&\sum_{\ell=0}^\infty \bra 0|e^{-\bt_2 T}|\ell\ket e^{-\lap\ell}
\bra\ell|e^{-\bt_1 T}|0\ket.
\end{aligned} 
\label{eq:2pt}
\end{equation}
Thus we find that the 2-point function is obtained by gluing two HH wavefunctions
$\bra\ell|e^{-\bt_{1,2} T}|0\ket$ together with the 
weight factor $e^{-\lap\ell}$ coming from the matter field.
Our result \eqref{eq:2pt} is schematically depicted as
\begin{equation}
\begin{aligned}
 \begin{tikzpicture}[scale=0.75]
\draw (0,0) circle [radius=2];
\draw[blue,thick] (-2,0)--(2,0); 
\draw (0,2) node [above]{$\bt_2$};
\draw (0,0) node [above]{$\ell$};
\draw (0,-2) node [below]{$\bt_1$};
\end{tikzpicture}
\end{aligned}\quad .
\label{eq:2pt-fig}
\end{equation}
The blue thick line in \eqref{eq:2pt-fig} represents
the matter chord. 
Recalling the definition of the state $|\ell\ket$ in
\eqref{eq:def-n} and replacing the dashed line of 
\eqref{eq:def-n} by the blue thick line, one can see that
the matter chord intersects the $H$-chord $\ell$ times 
and hence we get the factor $e^{-\lap\ell}$ in \eqref{eq:2pt}
\begin{equation}
\begin{aligned}
 \begin{tikzpicture}[scale=0.75]
\draw (0,-0.5)--(0,0.5);
\draw (0.5,-0.5)--(0.5,0.5);
\draw (1,-0.5)--(1,0.5);
\draw (1.5,-0.5)--(1.5,0.5);
\draw (2,-0.5)--(2,0.5);
\draw (2.5,-0.5)--(2.5,0.5);
\draw (3,-0.5)--(3,0.5);
\draw[blue, thick] (-0.5,0)--(3.5,0);
\draw (1.5,0.5) node [above] {$\overbrace{~\hskip23mm~}^{}$};
\draw (1.5,1) node [above] {$\ell~\text{chords}$};
\draw (4,0) node [right] {$=~~e^{-\lap\ell}$.};
\end{tikzpicture}
\end{aligned} 
\end{equation}
One can check that our result \eqref{eq:2pt} agrees with (C.5) in \cite{Berkooz:2018jqr} using our explicit form of the HH wavefunction in \eqref{eq:HH-sum}.

\subsection{4-point function}
Next consider the un-crossed 4-point function.
In principle, the un-crossed
4-point function can be computed
by using the relation \eqref{eq:matter-wick} twice. However,
we have to evaluate the remaining $\th$-integral, which is quite complicated.
Fortunately, this computation was already carried out in 
\cite{Berkooz:2018jqr} and the result is relatively simple; see
(C.7) in \cite{Berkooz:2018jqr}.
Our key observation is 
that (C.7) in \cite{Berkooz:2018jqr} is written in terms of the
HH wavefunctions $\bra k_i|e^{-\bt_i T}|0\ket$
\footnote{We believe that
$q^{k_{12}k_{34}}$ in (C.7) of \cite{Berkooz:2018jqr}
is a typo of $q^{k_{13}k_{24}}$.
Note also that in (C.7) of \cite{Berkooz:2018jqr} there is an extra factor $(-1)^{k_i}$ multiplying the HH wavefunction 
in \eqref{eq:HH-sum}. This sign factor can be removed by sending $\bt\to-\bt$ and using the relation $I_n(-z)=(-1)^n I_n(z)$, due to the symmetry under the sign flip
of the Hamiltonian $H\to -H$.
Alternatively, one can show that $\sum_i k_i$ is an even integer and hence
the sign factors are canceled in the final result.}
\begin{equation}
\begin{aligned}
 &\bra\tr e^{-\bt_4H}\wick{\c M_2e^{-\bt_3H}\c M_2} 
e^{-\bt_2 H}\wick {\c M_1e^{-\bt_1H}\c M_1}\ket_J\\
=&\sum_{\substack{k_{jl}=0\\1\leq j<l\leq4}}^\infty 
e^{-\lap_1 k_1}e^{-\lap_2k_3}q^{k_{13}k_{24}}
\prod_{i=1}^4\rt{\frac{(q;q)_{k_i}}{\prod_{i\in(jl)}(q;q)_{k_{jl}}}}
\bra k_i|e^{-\bt_i T}|0\ket,
\end{aligned} 
\label{eq:4pt}
\end{equation}
where 
\begin{equation}
\begin{aligned}
 k_i=\sum_{i\in (jl)}k_{jl}.
\end{aligned} 
\end{equation}
In other words, 
$k_i$ is obtained by summing all pairs of $(jl)$'s which contain $i$. 
More explicitly, $k_i$'s are written as
\begin{equation}
\begin{aligned}
 k_1&=k_{12}+k_{13}+k_{14},\quad
k_2=k_{12}+k_{23}+k_{24},\\
k_3&=k_{13}+k_{23}+k_{34},\quad
k_4=k_{14}+k_{24}+k_{34}. 
\end{aligned} 
\end{equation}
We notice that there is an extra square-root factor in  \eqref{eq:4pt}
multiplying the HH wavefunction. For instance, 
the factor for $i=1$ reads
\begin{equation}
\begin{aligned}
 \rt{\frac{(q;q)_{k_1}}{\prod_{1\in(jl)}(q;q)_{k_{jl}}}}
=\rt{\frac{(q;q)_{k_1}}{(q;q)_{k_{12}}(q;q)_{k_{13}}(q;q)_{k_{14}}}}
=\rt{
\begin{bmatrix}
 k_1\\k_{12},~k_{13},~k_{14}
\end{bmatrix}_q},
\end{aligned} 
\label{eq:prefac}
\end{equation}
where the $q$-multinomial coefficient is defined by
\begin{equation}
\begin{aligned}
 \begin{bmatrix}
 n\\n_1,~\cdots,~n_r
\end{bmatrix}_q =\frac{(q;q)_n}{\prod_{i=1}^r(q;q)_{n_i}},
\end{aligned} 
\end{equation} 
with $n=\sum_{i=1}^r n_i$. 
This result of 4-point function 
\eqref{eq:4pt} can be graphically represented as
\begin{equation}
\begin{aligned}
 \begin{tikzpicture}[scale=0.75]
\draw (0,0) circle [radius=3];
\draw[red,thick] (-2.6,1.5)--(-1,1); 
\draw[red,thick] (-1,1)--(1,1);
\draw[red,thick] (1,1)--(2.6,1.5); 
\draw[thick,dashed] (1,-1)--(1,1); 
\draw[thick,dashed] (-1,-1)--(-1,1); 
\draw[blue,thick] (-2.6,-1.5)--(-1,-1); 
\draw[blue,thick] (-1,-1)--(1,-1); 
\draw[blue,thick] (1,-1)--(2.6,-1.5);
\draw (0,-3) node [below]{$\bt_1$};
\draw (3,0) node [right]{$\bt_2$};
\draw (0,3) node [above]{$\bt_3$};
\draw (-3,0) node [left]{$\bt_4$}; 
\draw (-1,0) node [left]{$k_{24}$}; 
\draw (1,0) node [right]{$k_{24}$}; 
\draw (0,1) node [above]{$k_{13}$};
\draw (0,-1) node [below]{$k_{13}$};
\draw (1.7,-1.1) node [below]{$k_{12}$}; 
\draw (-1.6,-1.2) node [below]{$k_{14}$};
\draw (1.6,1.2) node [above]{$k_{23}$};
\draw (-1.6,1.2) node [above]{$k_{34}$};  
\end{tikzpicture}
\end{aligned} \quad,
\label{eq:4pt-diagram}
\end{equation}
where the blue and red thick lines represent 
the matter chords for $M_1$ and $M_2$, respectively.
As in the case of the 2-point function, 
the factors $e^{-\lap_1 k_1}$ and $e^{-\lap_2 k_3}$ in \eqref{eq:4pt}
come from the intersection between the matter chords and the $H$-chords.
Also, it is natural to interpret 
$q^{k_{13}k_{24}}$ in \eqref{eq:4pt}
as the factor counting the intersections of $H$-chords in the
middle rectangular part of the diagram
\eqref{eq:4pt-diagram}
\begin{equation}
\begin{aligned}
 \begin{tikzpicture}[scale=0.75]
\draw (1.2,0) node [right]{$\Biggl\}k_{24}~\text{chords}\quad 
=\quad q^{k_{13}k_{24}}$.}; 
\draw (0,1.2) node [above]{$\overbrace{~\hskip12mm~}^{}$};
\draw (0,1.7) node [above]{$k_{13}~\text{chords}$};
\draw[thick,dashed] (1,-1)--(1,1); 
\draw[thick,dashed] (-1,-1)--(-1,1);
\draw[blue,thick] (-1,-1)--(1,-1);
\draw[red,thick] (-1,1)--(1,1); 
\draw(-1.2,0.6)--(1.2,0.6);
\draw(-1.2,0.3)--(1.2,0.3);
\draw(-1.2,0)--(1.2,0);
\draw(-1.2,-0.3)--(1.2,-0.3);
\draw(-1.2,-0.6)--(1.2,-0.6); 
\draw(-0.8,-1.2)--(-0.8,1.2);
\draw(-0.6,-1.2)--(-0.6,1.2); 
\draw(-0.4,-1.2)--(-0.4,1.2); 
\draw(-0.2,-1.2)--(-0.2,1.2);
\draw(0,-1.2)--(0,1.2);  
\draw(0.8,-1.2)--(0.8,1.2);
\draw(0.6,-1.2)--(0.6,1.2); 
\draw(0.4,-1.2)--(0.4,1.2); 
\draw(0.2,-1.2)--(0.2,1.2);
\end{tikzpicture}
\end{aligned}
\label{eq:rectangle}
\end{equation}
We do not have a clear understanding of the $q$-multinomial factor
\eqref{eq:prefac} multiplying the HH wavefunction.
Perhaps, this accounts for some combinatorics of
dividing the $k_1$ chords into three groups with chord numbers
$(k_{12},k_{13},k_{14})$. It would be interesting to understand this 
factor better.

We also recognize that the un-crossed $2n$-point function in
(C.10) of \cite{Berkooz:2018jqr}
is written in terms of the HH wavefunctions
$\bra k_i|e^{-\bt_i T}|0\ket$
\begin{equation}
\begin{aligned}
 &\bra\tr e^{-\bt_{2n}H}\wick{\c M_n e^{-\bt_{2n-1}H}\c M_n}\cdots
\wick {\c M_1e^{-\bt_1H}\c M_1}\ket_J\\
=&
\sum_{\substack{k_{jl}=0\\1\leq j<l\leq 2n}}^\infty q^{B(k_{jl})}
\prod_{m=1}^n e^{-\lap_m k_{2m-1}}\prod_{i=1}^{2n}
\rt{\frac{(q;q)_{k_i}}{\prod_{i\in(jl)}(q;q)_{k_{jl}}}}
\bra k_i|e^{-\bt_i T}|0\ket,
\end{aligned}
\label{eq:2n-pt} 
\end{equation}
where $k_i=\sum_{i\in (jl)}k_{jl}$ and
\begin{equation}
\begin{aligned}
 B(k_{jl})=\sum_{1\leq j_1<j_2<j_3<j_4\leq 2n}
k_{j_1j_3}k_{j_2j_4}.
\end{aligned} 
\end{equation}
Again, we expect that \eqref{eq:2n-pt} has a natural
interpretation in the bulk gravitational picture 
as a certain generalization of
 the diagram for the 4-point function \eqref{eq:4pt-diagram}.

\section{Probability distribution of $\ell$}\label{sec:prob}
If we naively apply the probability interpretation of quantum mechanics
to the ``wavefunction of the universe'' $\bra\ell|e^{-\bt T}|0\ket$,
then its absolute value squared
$|\bra\ell|e^{-\bt T}|0\ket|^2$ can be thought of as the probability
distribution of $\ell$.
It seems more natural to normalize $|\bra\ell|e^{-\bt T}|0\ket|^2$
by the partition function $Z(2\bt)=\bra0|e^{-2\bt T}|0\ket$
\begin{equation}
\begin{aligned}
 P_\ell(\bt,q):=\frac{\bra0|e^{-\bt T}|\ell\ket\bra\ell|e^{-\bt T}|0\ket}{\bra0|e^{-2\bt T}|0\ket}
\end{aligned} 
\end{equation}
in such a way that the total probability becomes unity
\begin{equation}
\begin{aligned}
 \sum_{\ell=0}^\infty P_\ell(\bt,q)=1.
\end{aligned} 
\end{equation}
Here we assumed $\bt\in\mathbb{R}$.
Using this definition of probability, one can consider various quantities 
associated with this distribution. For instance
we can consider the expectation value of
$\ell$
\begin{equation}
\begin{aligned}
 \b{\ell}(\bt,q):=\sum_{\ell=0}^\infty\ell P_\ell(\bt,q).
\end{aligned} 
\end{equation} 
When $q=1$, the distribution of $\ell$ becomes the Poisson distribution
\begin{equation}
\begin{aligned}
 P_\ell(\bt,1)=\frac{\bt^{2\ell}}{\ell!}e^{-\bt^2},
\end{aligned} 
\end{equation} 
since the $q$-deformed oscillator reduces to the ordinary oscillator
$a,a^\dagger$ with $[a,a^\dagger]=1$,
and the Hartle-Hawking state $e^{-\bt T}|0\ket$ 
becomes the coherent state $e^{\bt(a+a^\dag)}|0\ket$ of ordinary
harmonic oscillator.
In this case, the expectation value of $\ell$ is given by
\begin{equation}
\begin{aligned}
 \b{\ell}(\bt,1)=\bt^2.
\end{aligned} 
\label{eq:vev-poisson}
\end{equation}
For $0<q<1$, we do not know the analytic form of $\b{\ell}(\bt,q)$
as a function of $\bt$ and $q$. Instead, we can compute
it numerically by truncating the sum
over $r$ in \eqref{eq:HH-sum} up to some cut-off $r\leq r_{\text{cut}}$.

\begin{figure}[t]
\centering
\subcaptionbox{$\bt=1,q=0.1$\label{sfig:P1}}{\includegraphics
[width=0.45\linewidth]{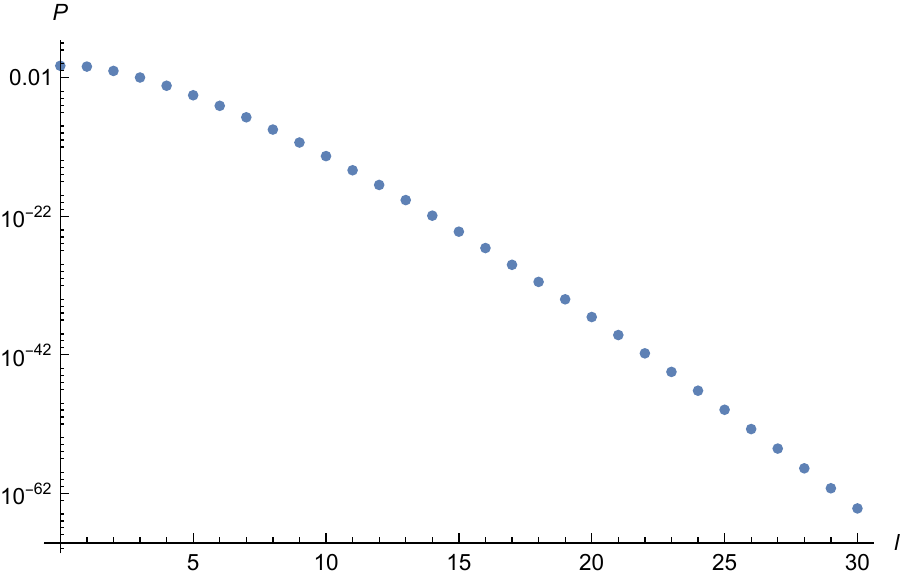}}
\hskip5mm
\subcaptionbox{$\bt=10,q=0.1$\label{sfig:P2}}{\includegraphics
[width=0.45\linewidth]{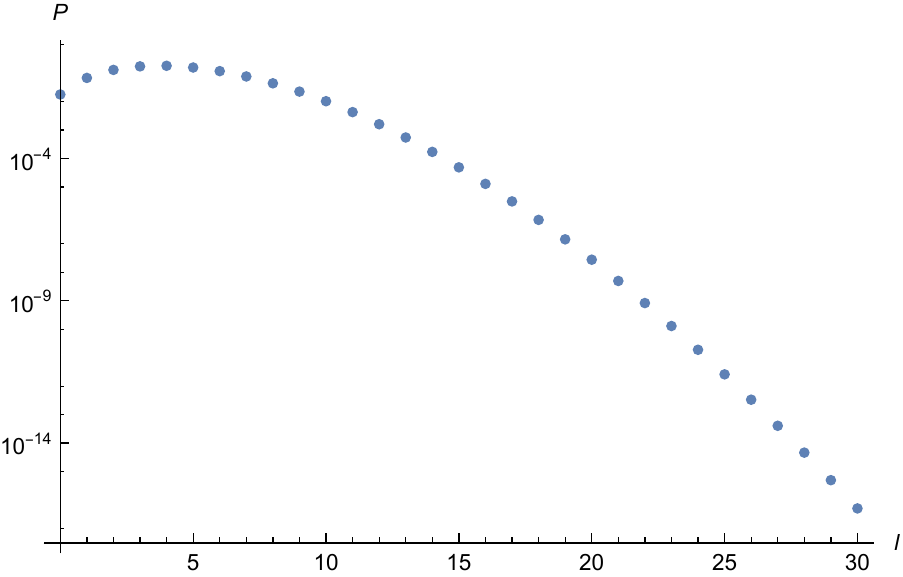}}\\
\vskip2mm
\centering
\subcaptionbox{$\bt=1,q=0.8$\label{sfig:P3}}{\includegraphics
[width=0.45\linewidth]{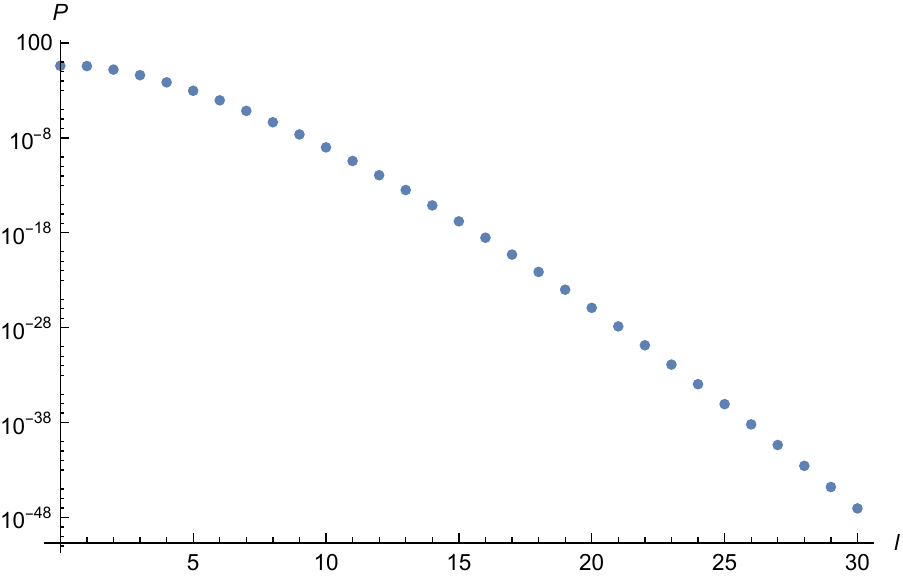}}
\hskip5mm
\subcaptionbox{$\bt=10,q=0.8$\label{sfig:P4}}{\includegraphics
[width=0.45\linewidth]{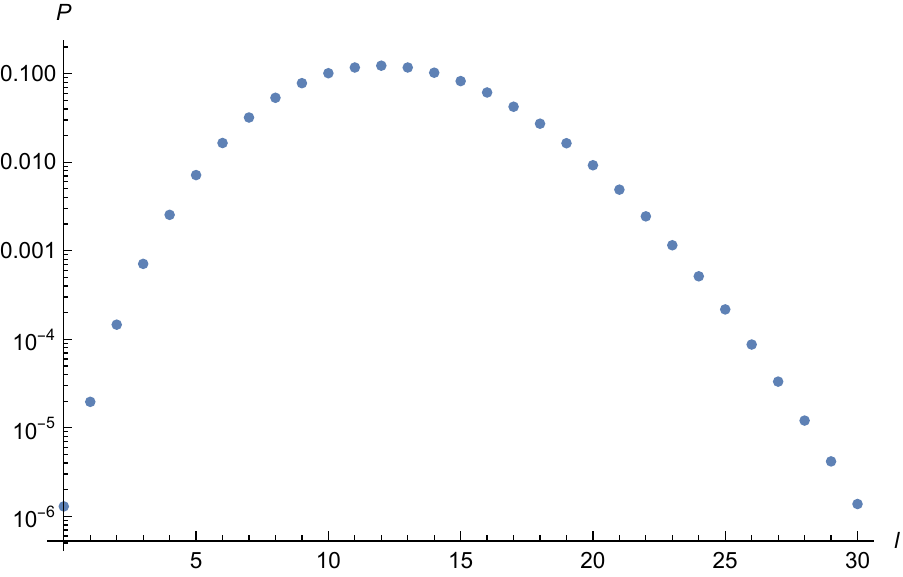}}
  \caption{
Plot 
of $P_\ell(\bt,q)$ against $\ell$ for various values of $\bt$ and $q$.
}
  \label{fig:P}
\end{figure}

\begin{figure}[t]
\centering
\subcaptionbox{$q=0.1$\label{sfig:l1}}{\includegraphics
[width=0.45\linewidth]{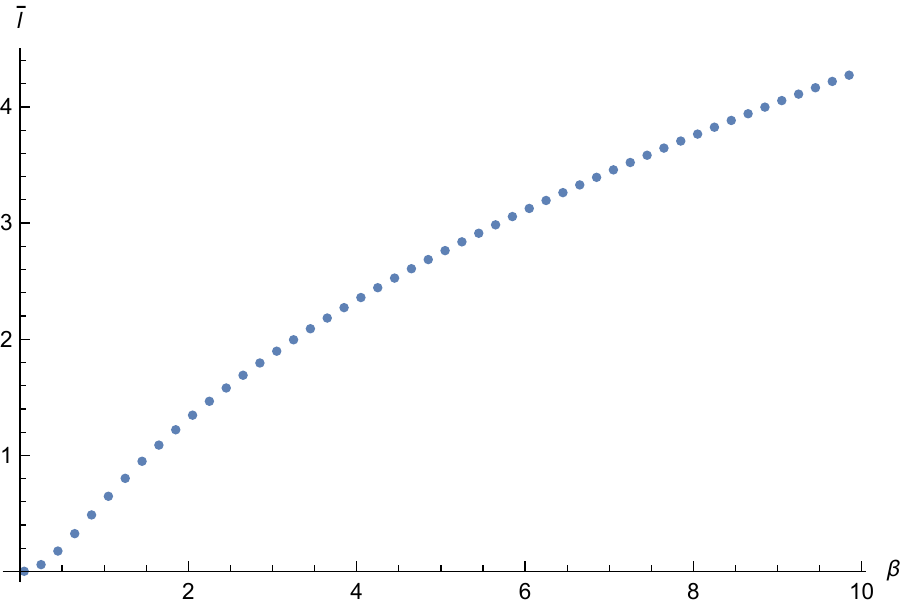}}
\hskip5mm
\subcaptionbox{$q=0.8$\label{sfig:l2}}{\includegraphics
[width=0.45\linewidth]{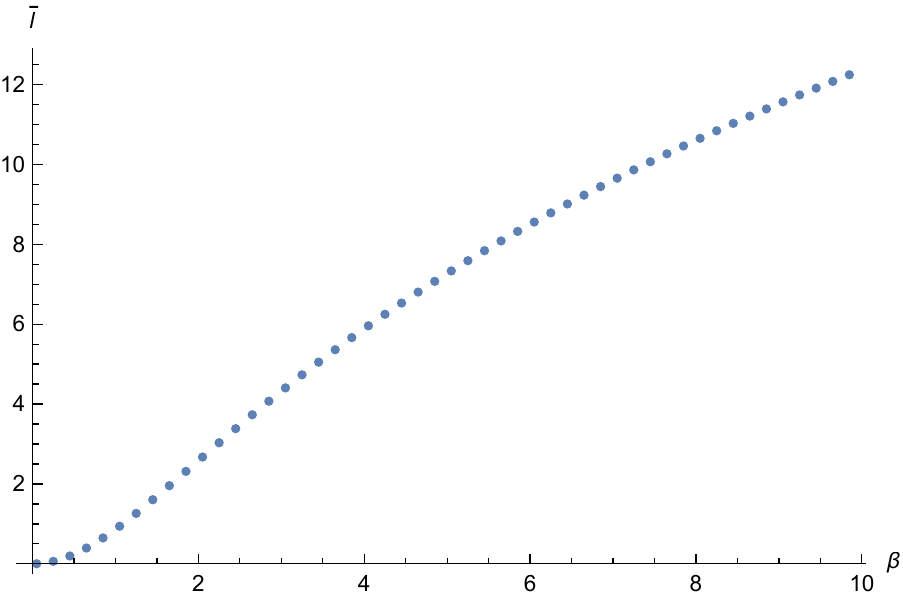}}
  \caption{
Plot 
of $\b{\ell}(\bt,q)$ against $\bt$ with fixed $q$.
}
  \label{fig:vev-ell}
\end{figure}

In Figure~\ref{fig:P}, we show the plot of $P_\ell(\bt,q)$
for various values of $\bt$ and $q$.
When $\bt$ is small (or at high temperature), $P_\ell(\bt,q)$ is peaked around 
$\ell=0$ and the large value of $\ell$ is highly suppressed.
On the other hand, 
when $\bt$ becomes large (or at low temperature)
$P_\ell(\bt,q)$ has a maximum at a non-zero value of $\ell$.
In Figure~\ref{fig:vev-ell}, we show the plot of the expectation value
of $\ell$ as a function of $\bt$. One can see that
$\b{\ell}(\bt,q)$ is a monotonically increasing function of $\bt$,
but the growing rate for $q<1$ is much slower than the $q=1$ case
\eqref{eq:vev-poisson}.

The probability distribution $P_\ell(\bt,q)$ appears in the 2-point function
\eqref{eq:2pt} when $\bt_1=\bt_2=\bt$
\begin{equation}
\begin{aligned}
 \bra\tr e^{-\bt H}\wick{\c M e^{-\bt H}\c M}\ket_J=
Z(2\bt)\sum_{\ell=0}^\infty e^{-\lap\ell}P_\ell(\bt,q).
\end{aligned} 
\end{equation} 
As a crude approximation, we can replace $\ell$ by its mean
$\b{\ell}(\bt,q)$
\begin{equation}
\begin{aligned}
 \bra\tr e^{-\bt H}\wick{\c M e^{-\bt H}\c M}\ket_J\approx Z(2\bt)e^{-\lap
\b{\ell}(\bt,q)}.
\end{aligned} 
\end{equation}
It would be interesting to understand the validity of this approximation.

We should stress that $\ell$ is a dynamical variable
and we have to sum over $\ell$ in order to obtain a physical quantity.
For instance, as shown in \eqref{eq:2pt} 
the physical 2-point function of matter field is obtained
only after we sum over $\ell$.

\section{Conclusions and outlook}\label{sec:conclusion}
In this paper, we have computed the HH wavefunction
$\bra\ell|e^{-\bt T}|0\ket$ in the DSSYK model explicitly,
and we observed that the known result of un-crossed correlator of matter
fields in \cite{Berkooz:2018jqr} 
is obtained by gluing the HH wavefunctions
with an appropriate weight.
This observation suggests the bulk gravitational
picture for the
2-point function in \eqref{eq:2pt-fig} and
for the 4-point function in \eqref{eq:4pt-diagram}.
Interestingly, the basic building block of the bulk spacetime is the chord
and the spacetime can be thought of as fabric
made of the chords (see e.g. \eqref{eq:rectangle}).
This picture is vaguely reminiscent of the random tensor network 
\cite{Hayden:2016cfa} or 
the bit threads \cite{Freedman:2016zud}.
As emphasized in \cite{Berkooz:2018qkz,Berkooz:2018jqr,Lin:2022rbf},
in the DSSYK model
the geodesic length is replaced by the number of chords $\ell$ which is naturally
quantized. 
It would be very interesting to extract more information of the bulk spacetime
from the DSSYK model.

There are many interesting open questions.
The crossed 4-point function in DSSYK
is constructed in \cite{Berkooz:2018jqr}
and it involves a complicated $R$-matrix originated from the
underlying quantum group symmetry.
It would be nice if we can extract the bulk spacetime picture 
of the crossed 4-point function. In particular 
we would like to understand the relation 
between the crossed 4-point function and 
the HH wavefunction as we did for the un-crossed case.
It is also interesting to go beyond the
strict large $N$ limit and compute the $1/N$ corrections.
In the strict large $N$ limit \eqref{eq:scaling}, we only see the bulk
spacetime with disk topology.
It would be interesting to study the higher genus corrections to
the DSSYK model and develop a technique to compute the
higher genus corrections systematically. As a first step,
it would be interesting to compute the DSSYK analogue of the 
``trumpet partition function'', 
which played an important role in the 
matrix model of JT gravity \cite{Saad:2019lba}.
We leave this as a very interesting future problem.

\acknowledgments
This work was supported
in part by JSPS Grant-in-Aid for Transformative Research Areas (A) 
``Extreme Universe'' No. 21H05187 and JSPS KAKENHI Grant No. 22K03594.

\appendix

\bibliography{paper}

\providecommand{\href}[2]{#2}\begingroup\raggedright\begin{thebibliography}{10}

\bibitem{kitaev1}
A.~Kitaev, ``A simple model of quantum holography (part 1),''.
  \url{https://online.kitp.ucsb.edu/online/entangled15/kitaev/}.

\bibitem{kitaev2}
A.~Kitaev, ``A simple model of quantum holography (part 2),''.
  \url{https://online.kitp.ucsb.edu/online/entangled15/kitaev2/}.

\bibitem{Sachdev1993}
S.~Sachdev and J.~Ye, ``Gapless spin-fluid ground state in a random quantum
  heisenberg magnet,''
  \href{http://dx.doi.org/10.1103/physrevlett.70.3339}{{\em Phys. Rev. Lett.}
  {\bfseries 70} no.~21, (1993) 3339--3342},
  \href{http://arxiv.org/abs/cond-mat/9212030}{{\ttfamily
  arXiv:cond-mat/9212030}}.

\bibitem{Maldacena:2016hyu}
J.~Maldacena and D.~Stanford, ``{Remarks on the Sachdev-Ye-Kitaev model},''
  \href{http://dx.doi.org/10.1103/PhysRevD.94.106002}{{\em Phys. Rev. D}
  {\bfseries 94} no.~10, (2016) 106002},
  \href{http://arxiv.org/abs/1604.07818}{{\ttfamily arXiv:1604.07818
  [hep-th]}}.

\bibitem{Polchinski:2016xgd}
J.~Polchinski and V.~Rosenhaus, ``{The Spectrum in the Sachdev-Ye-Kitaev
  Model},'' \href{http://dx.doi.org/10.1007/JHEP04(2016)001}{{\em JHEP}
  {\bfseries 04} (2016) 001}, \href{http://arxiv.org/abs/1601.06768}{{\ttfamily
  arXiv:1601.06768 [hep-th]}}.

\bibitem{Jackiw:1984je}
R.~Jackiw, ``{Lower Dimensional Gravity},''
\href{http://dx.doi.org/10.1016/0550-3213(85)90448-1}{{\em Nucl. Phys.}
  {\bfseries B252} (1985) 343--356}.

\bibitem{Teitelboim:1983ux}
C.~Teitelboim, ``{Gravitation and Hamiltonian Structure in Two Space-Time
  Dimensions},''
\href{http://dx.doi.org/10.1016/0370-2693(83)90012-6}{{\em Phys. Lett.}
  {\bfseries 126B} (1983) 41--45}.

\bibitem{Cotler:2016fpe}
J.~S. Cotler, G.~Gur-Ari, M.~Hanada, J.~Polchinski, P.~Saad, S.~H. Shenker,
  D.~Stanford, A.~Streicher, and M.~Tezuka, ``{Black Holes and Random
  Matrices},'' \href{http://dx.doi.org/10.1007/JHEP05(2017)118}{{\em JHEP}
  {\bfseries 05} (2017) 118}, \href{http://arxiv.org/abs/1611.04650}{{\ttfamily
  arXiv:1611.04650 [hep-th]}}. [Erratum: JHEP 09, 002 (2018)].

\bibitem{Berkooz:2018qkz}
M.~Berkooz, P.~Narayan, and J.~Simon, ``{Chord diagrams, exact correlators in
  spin glasses and black hole bulk reconstruction},''
  \href{http://dx.doi.org/10.1007/JHEP08(2018)192}{{\em JHEP} {\bfseries 08}
  (2018) 192}, \href{http://arxiv.org/abs/1806.04380}{{\ttfamily
  arXiv:1806.04380 [hep-th]}}.

\bibitem{Berkooz:2018jqr}
M.~Berkooz, M.~Isachenkov, V.~Narovlansky, and G.~Torrents, ``{Towards a full
  solution of the large N double-scaled SYK model},''
  \href{http://dx.doi.org/10.1007/JHEP03(2019)079}{{\em JHEP} {\bfseries 03}
  (2019) 079}, \href{http://arxiv.org/abs/1811.02584}{{\ttfamily
  arXiv:1811.02584 [hep-th]}}.

\bibitem{Lin:2022rbf}
H.~W. Lin, ``{The bulk Hilbert space of double scaled SYK},''
  \href{http://dx.doi.org/10.1007/JHEP11(2022)060}{{\em JHEP} {\bfseries 11}
  (2022) 060}, \href{http://arxiv.org/abs/2208.07032}{{\ttfamily
  arXiv:2208.07032 [hep-th]}}.

\bibitem{Hartle:1983ai}
J.~B. Hartle and S.~W. Hawking, ``{Wave Function of the Universe},''
  \href{http://dx.doi.org/10.1103/PhysRevD.28.2960}{{\em Phys. Rev. D}
  {\bfseries 28} (1983) 2960--2975}.

\bibitem{Saad:2019pqd}
P.~Saad, ``{Late Time Correlation Functions, Baby Universes, and ETH in JT
  Gravity},'' \href{http://arxiv.org/abs/1910.10311}{{\ttfamily
  arXiv:1910.10311 [hep-th]}}.

\bibitem{Hayden:2016cfa}
P.~Hayden, S.~Nezami, X.-L. Qi, N.~Thomas, M.~Walter, and Z.~Yang,
  ``{Holographic duality from random tensor networks},''
  \href{http://dx.doi.org/10.1007/JHEP11(2016)009}{{\em JHEP} {\bfseries 11}
  (2016) 009}, \href{http://arxiv.org/abs/1601.01694}{{\ttfamily
  arXiv:1601.01694 [hep-th]}}.

\bibitem{Freedman:2016zud}
M.~Freedman and M.~Headrick, ``{Bit threads and holographic entanglement},''
  \href{http://dx.doi.org/10.1007/s00220-016-2796-3}{{\em Commun. Math. Phys.}
  {\bfseries 352} no.~1, (2017) 407--438},
  \href{http://arxiv.org/abs/1604.00354}{{\ttfamily arXiv:1604.00354
  [hep-th]}}.

\bibitem{Saad:2019lba}
P.~Saad, S.~H. Shenker, and D.~Stanford, ``{JT gravity as a matrix integral},''
  \href{http://arxiv.org/abs/1903.11115}{{\ttfamily arXiv:1903.11115
  [hep-th]}}.

\end{thebibliography}\endgroup
\bibliographystyle{utphys}

\end{document}